# The connection between zero chromaticity and long in-plane polarization lifetime In a magnetic storage ring

G. Guidoboni,[1] E.J. Stephenson,[2] A Wrońska,[3] Z. Bagdasarian,[4,5] J. Bsaisou,[6] S. Chekmenev,[7] S. Dymov,[4,8] D. Eversmann,[7] M. Gaisser,[9] R. Gebel,[4] V. Hejny,[4] N. Hempelmann,[4] F. Hinder,[7] A. Kacharava,[4] I. Keshelashvili,[4] P. Kulessa,[10] P. Lenisa,[1] A. Lehrach,[4,11] B. Lorentz,[4] P. Maanen,[7] R. Maier,[4,11] D. Mchedlishvili,[4,5] S. Mey,[4] A. Nass,[4] A. Pesce,[1] Y. Orlov,[12] J. Pretz,[7,11] D. Prasuhn,[4] F. Rathmann,[4] M. Rosenthal,[4] A. Saleev,[4,13] Y.K. Semertzidis,[9] Y. Senichev,[4] V. Shmakova,[4,14] H. Stockhorst,[4] H. Ströher,[4] R. Talman,[12] P. Thörngren-Engblom,[1,15] F. Trinkel,[4] Yu. Valdau,[4] C. Weidemann,[4] P. Wüstner,[16] and D. Zyusin[4]

[1]*University of Ferrara and INFN, 44122 Ferrara, Italy*
[2]*Indiana University Center for Spacetime Symmetries, Bloomington, 47405 Indiana, USA*
[3]*Institute of Physics, Jagellonian University, 30348 Cracow, Poland*
[4]*Institut für Kernphysik, Forschungszentrum Jülich, 52425 Jülich, Germany*
[5]*High Energy Physics Institute, Tbilisi State University, 0186 Tbilisi, Georgia*
[6]*Institute for Advanced Simulation, Forschungszentrum Jülich, 52425 Jülich, Germany*
[7]*III. Physikalishes Institut B, RWTH Aachen University, 52056 Aachen, Germany*
[8]*Laboratory of Nuclear Problems, Joint Institute for Nuclear Research, 141980 Dubna, Russia*
[9]*Center for Axion and Precision Physics research, Institute for Basic Science, 291 Daehak-ro, Yuseong-gu, Daejeon 305-701, Republic of Korea*
[10]*Institute of Nuclear Physics PAN, 31-342 Krakow, Poland*
[11]*JARA-FAME (Forces and Matter Experiments), Forschungszentrum Jülich and RWTH Aachen University, Germany*
[12]*Cornell University, Ithaca, New York 14850, USA*
[13]*Samara National Research University, 443086 Samara, Russia*
[14]*Institute for Theoretical and Experimental Physics, 117259 Moscow, Russia*
[15]*Department of Physics, Royal Institute of Technology, 10691 Stockholm, Sweden*
[16]*Zentralinstitut für Engineering, Elektronik, und Analytik, Forschungszentrum Jülich, 52425 Jülich, Germany*

ABSTRACT: In this paper, we demonstrate the connection between a magnetic storage ring with additional sextupole fields set so that the *x* and *y* chromaticities vanish and the maximizing of the lifetime of in-plane polarization (IPP) for a 0.97-GeV/*c* deuteron beam. The IPP magnitude was measured by continuously monitoring the down-up scattering asymmetry (sensitive to sideways polarization) in an in-beam, carbon-target polarimeter and unfolding the precession of the IPP due to the magnetic anomaly of the deuteron. The optimum operating conditions for a long IPP lifetime were made by scanning the field of the storage ring sextupole magnet families while observing the rate of IPP loss during storage of the beam. The beam was bunched and electron cooled. The IPP losses appear to arise from the change of the orbit circumference, and consequently the particle speed and spin tune, due to the transverse betatron oscillations of individual particles in the beam. The effects of these changes are canceled by an appropriate sextupole field setting.

# I. INTRODUCTION

A recent letter [1] reports the achievement of a 1000-second lifetime for the in-plane polarization (IPP) of a spin-polarized deuteron beam circulating at 0.97-GeV/c in the COSY storage ring [2]. This lifetime was made possible through a combination of beam bunching, electron cooling, trimming of sextupole field components, and limiting the beam current ($< 10^9$ deuterons/fill). This demonstration fulfills a requirement [3], albeit at a lower circulating beam current, for the use of a storage ring to search for an intrinsic electric dipole moment (EDM) on the nuclei in a circulating charged-particle beam. This paper describes in more detail the connection noted in Fig. 3 of that letter between the setting of the COSY sextupole trim fields for long IPP lifetime and the vanishing of the $x$ and $y$ chromaticities in the ring. The COSY storage ring was operated at horizontal and vertical betatron tunes near 3.6 (but not equal to each other) and well away from machine resonances [4]. The features of an EDM storage ring experiment have been reviewed elsewhere [1,3,5].

The precession of the beam's IPP in a storage ring is governed by the interaction between the particles' magnetic anomaly and any vertical (perpendicular to the ring plane) magnetic field. The precession rate is given by the product of the beam revolution frequency $f_{\text{rev}}$ and the spin tune $v_S = G\gamma$ where $G$ is the deuteron's magnetic anomaly and $\gamma$ is the relativistic factor. Any variation in the speeds among the beam particles creates a distribution in $\gamma$ that leads to a spread in individual particle spin directions and eventual depolarization of the beam. With a coasting beam distributed around the ring, the leading first-order contribution to $v_S$ variations comes from the momentum spread of the beam, which is typically in the vicinity of $10^{-4}$ of the nominal momentum for the usual sorts of beam injection schemes. This leads to an IPP lifetime of tens of milliseconds for $f_{\text{rev}} \approx 10^6$ s$^{-1}$. This effect can be removed on average by bunching the beam, thus forcing long-term isochronicity on all of the beam particles. Variations in individual particle momenta still exist, but the rf cavity now traps all particles within the bunch where they oscillate in synchrotron orbits of different sizes. Since each particle's momentum now oscillates about the average value, depolarization is significantly reduced.

Second-order contributions to the spin tune spread may come from oscillations of the particles about the nominal orbit through either betatron (transverse) or synchrotron (longitudinal) motion. The transverse motion increases the length of the path that the particles must travel as they go around the ring. In combination with beam bunching, this forces those particles with large oscillation amplitudes to travel with a larger speed, thus increasing $\gamma$ and generating depolarization.

If the oscillations in the longitudinal direction are of small amplitude compared to the ring circumference and the rf cavity is driven by a sinusoidal voltage, then the motion closely approximates a simple harmonic system. If the effect of the speed increase as the particle moves forward through the beam bunch is balanced by a speed decrease while going in the backward direction, then there is no net effect on average and these speed changes do not contribute to a growing depolarization. Typically, the frequency of this oscillation is about four orders of magnitude smaller than the revolution frequency while betatron oscillations in a strong focusing system have frequencies that are larger than the revolution frequency by a factor of a few.

For large synchrotron oscillations at higher order, an additional quadratic dependence of $\gamma$ on speed appears and this term may result in a shift of the average spin tune from the reference orbit value. The size of the spin tune shift depends on the second-order momentum compaction factor [6], a property of the ring lattice. This effect is hard to measure for the COSY lattice at the lower beam momenta relevant for this experiment due to limitations in the acceptance of the ring for momentum changes and the requirement to measure a quadratic term. This and other evidence suggest that these higher-order effects of synchrotron motion are small enough that we can neglect them for the studies described here.

The second-order effects from transverse orbit oscillations may be reduced by applying electron cooling to the beam, which reduces the spread of particle orbits in phase space. Under these conditions, the IPP lifetime may reach seconds instead of tens of milliseconds.

A better way to manage second-order depolarization from transverse oscillations is to cancel the effect entirely rather than just reduce the beam's phase space size. Sextupole fields in conjunction with quadrupole focusing fields can shift the center point of the transverse oscillation pattern. The amount of the shift depends quadratically on the size of the transverse oscillation. Such adjustments made in the arcs of the COSY ring can result in shorter orbits when the shift is toward the center of the ring. In combination with the path lengthening created by the oscillations in the first place, a proper sextupole strength can match these two effects, making the modified orbits the same length as unperturbed orbits with no transverse oscillation. Since the isochronicity of the orbit is maintained by the rf cavity, all orbits will have the same average spin tune. The COSY ring was built with sextupole correction magnets located where there are large horizontal and vertical beta functions as well as a large horizontal dispersion. The available sextupole fields appear to be sufficient to create the needed cancellations. So the studies of the beam properties undertaken in this paper explores how to measure and then correct the effects of depolarization in the presence of a distribution of transverse oscillation amplitudes.

Section II contains a brief summary of transverse oscillations and their connection to chromaticity. In Section III we turn to the explanation of the experiment, beginning with a review of the time-marking system used to measure the IPP [7] and a description of the context provided by the map of chromaticity values for the data taking plan. The results are shown in Section IV along with the constraints on their interpretation. This section also compares the best IPP lifetimes with the chromaticity values. Conclusions are presented in Section V.

## II.   TRANSVERSE OSCILLATIONS, CHROMATICITY, AND SEXTUPOLE FIELD ADJUSTMENTS

In a strong focusing machine, particles that are undergoing transverse oscillations will see a restoring force toward the reference orbit that is strong enough to generate a few oscillations for one revolution of the beam around the machine. The number of oscillations per revolution is known as the betatron tune, $v$. The horizontal ($x$) and vertical ($y$) focusing histories are different and not strongly coupled to each other. So it makes sense to define betatron tunes for both directions, either $v_x$ or $v_y$. In general, these tunes depend on the particle momentum $p$. The linear coefficient in this relationship is called the chromaticity, $\xi_{x,y} = \partial v_{x,y}/\partial(\Delta p/p)$. Since an increase in the beam momentum reduces the time that is available for oscillations during one revolution, and since the restoring force from the quadrupoles remains roughly the same for all momenta, chromaticities in lattices without other alterations tend to be negative. This is called the natural chromaticity $\xi_{nat}$. This result may be altered, as we will see later, by the addition of sextupole fields, some of which may arise in standard components such as the ring dipoles.

The natural chromaticity for one particle may be expressed in terms of the Courant-Snyder parametrization of betatron oscillations in a velocity vs. position phase space with the symbols $\alpha_{cs}$, $\beta_{cs}$ and $\gamma_{cs}$ (see Chap. 2 of Ref. [8]). In the case of vertical oscillations, for example, this leads to

$$4\pi\xi_{nat} = \oint K\beta_{cs}\,ds = \oint (\gamma_{cs} + \alpha_{cs}')\,ds = \oint \gamma_{cs}\,ds \tag{1}$$

where $K$ is the focusing function and $ds$ is a differential element of the beam path around the ring. A common substitution, $K\beta_{cs} = \gamma_{cs} + \alpha_{cs}'$, is used. The derivative of $\alpha_{cs}$ is with respect to $s$ and its integral around the ring vanishes for one complete turn. For horizontal oscillations, Eq. (1) would begin with a minus (−) sign.

Betatron oscillations increase the path length or circumference around the ring as

$$C = \oint \sqrt{1 + y'^2}\,ds \approx C_0 + \frac{1}{2}\oint y'^2\,ds \tag{2}$$

where $C_0$ is the circumference of the reference orbit and $y'$ is the slope of the particle path with respect to $s$. The fractional change in the circumference is given by

$$\frac{\Delta C}{C_0} = \frac{1}{2C_0}\oint y'^2\,ds = \frac{J}{C_0}\oint \gamma_{cs}\,ds = 4\pi\frac{J}{C_0}\xi_{nat} \tag{3}$$

where $J$ is the action (see Chap. 2.1.1 of Ref. [9] for the substitution). The last substitution comes from Eq. (1). This result is similar to Ref. [10] for an ensemble of particles.

Minimizing the depolarization means that the path length changes must become small or zero for particles away from the reference orbit. This means that $\Delta C$ must become zero. This can only be accomplished in Eq. (3) by including an additional contribution to the chromaticity from sextupole fields, as given by

$$\frac{\Delta C}{C_0} = 4\pi\frac{J}{C_0}\xi_{nat} - \frac{J}{C_0}D\beta\frac{B''\ell}{(B\rho)} \tag{4}$$

where $D$ is the dispersion, $B''$ is the curvature of the sextupole field, $\ell$ is the sextupole magnet length, and $(B\rho)$ is the rigidity of the ring (see Chap. 2.1.1 of Ref. [9]). Making $\Delta C = 0$ means that the total chromaticity vanishes.

It has been argued independently along these same lines that the simultaneous appearance of zero chromaticity and a long IPP lifetime should occur [11]. This property was used in the preparation of the electron and positron beams for a comparison of their magnetic moments [12,13].

The appearance of the action $J$ in both terms of Eq. (4) means that the strength of the sextupole, adjusted by changing $B''$, may be used to make the variations in the circumference zero for all values of $J$ or oscillation sizes, at least to this order. More generally, we can state the requirement for sextupole strengths as follows

$$0 = |A + a_1 I_S + a_2 I_L + a_3 I_G|\langle x'^2\rangle + |B + b_1 I_S + b_2 I_L + b_3 I_G|\langle y'^2\rangle + O\langle(\Delta p/p)^2\rangle \tag{5}$$

where *A* and *B* represent the contributions to the natural chromaticity, or the chromaticity with no adjustable sextupole strength applied. The sextupole currents, $I_S$, $I_L$, and $I_G$ are the currents applied to the three families of sextupoles located in the arcs of the COSY ring. These families are located at places of large Courant-Snyder $\beta_{x,cs}$, $\beta_{y,cs}$ and dispersion *D* respectively. The subscripts *S*, *L* and *G* refer to designations of the sextupole families (MXS, MXL, and MXG) used in the magnet naming scheme for COSY. These beam properties define the sensitivity to the sextupole strength described by the $a_i$ and $b_i$ coefficients. But the currents are the same in both of the major terms of Eq. (5), and this means that any solution must be found for both terms together. It may also be the case that there is an additional dependence, but through synchrotron oscillations rather than betatron that adds an additional term that depends on $\langle (\Delta p/p)^2 \rangle$. Such a term arises only as the synchrotron motion departs from simple harmonic, as it eventually must in the sinusoidal potential provided by the ring rf bunching cavity.

       The requirement as sextupole fields are adjusted is that the IPP lifetime improves. In practical usage, we often tracked the reciprocal of the IPP lifetime and looked for a setting where the reciprocal went to zero (or close to zero). This will come up again in Section IV where lifetime measurements are explained in more detail. In an independent check, model calculations of betatron-induced depolarization made using the no-lattice spin transport model of Benati [14] showed that the reciprocal of the polarization lifetime varies linearly with the size of the betatron oscillations as represented by the square of the transverse velocity or particle path slope shown in Eq. (1). Thus the behavior of the reciprocal near zero chromaticity should be linear with sextupole strength.

       These arguments provided a framework to search for long IPP lifetimes with the assumption that the best conditions under which to find the optimum lifetime would occur when $\xi_x = 0$ and $\xi_y = 0$ at the same time. This may be understood in a simple way by the statement that all particle paths have the same length regardless of the magnitude of any transverse (betatron) oscillation that may be present.

### III.   EVENT TIME MARKING, CHROMATICITY MEASUREMENT, AND RUNNING PLAN

       A summary of the experimental method was presented by Guidoboni~\cite{Guido2016}. The beam polarization was continuously monitored using scattering of beam deuterons from a thick carbon target into a set of polarimeter scintillators used previously by the EDDA experiment~\cite{Weise,Albers,Bisp}. Each polarimeter event was marked with a clock time from which it was possible to unfold the anomalous precession based on the total time that had elapsed since the beginning of data acquisition on a particular beam store and an estimate of the spin tune. From this, we assigned an IPP direction or phase to each event. In each of a series of time bins during the store, the polarimeter events were sorted by direction and the asymmetry computed for each direction bin from a comparison of the down and up counting rates assigned to that direction bin. A fit of a sinusoidal curve to these asymmetries would then give the magnitude of the IPP and the direction, or phase, of the polarization at the beginning of the time bin.

       In this analysis, the spin tune must be accurate to about one part in $10^6$ or else the accumulated difference between the calculated and actual phases in even a one-second time bin is

enough to reduce the apparent IPP magnitude. To remedy this, the spin tune in the analysis was scanned over a narrow range near the starting value and a search was conducted for the spin tune with the maximum IPP. This allowed each time interval to have its own spin tune, but in practice this led to an unstable series of phase values. In a second pass through the data, the spin tune was fixed at an average value among all the time bins (sometimes with a smooth trend added) and the analysis repeated. With this constraint, the new set of phases had a smooth time dependence so long as the IPP magnitude was not close to zero [7].

During a series of runs recorded during a 2015 experiment, the injected beam going into COSY was equal to or less than $10^9$ deuterons/fill. This led to measurements of the IPP as a function of the time during the store as shown in Fig. 1 of Ref. [1]. However, the more detailed comparison of sextupole settings for a good IPP lifetime with values of the chromaticity, as shown in Fig. 3 of Ref. [1], was made with higher beam currents (up to an order of magnitude) in a longer study completed in 2014. It is this longer study that we describe in this paper.

At the start of the experiment, the chromaticities of the COSY ring were measured for a large grid of sextupole settings. The beam was prepared in coasting mode (no bunching) with electron cooling on, and small velocity changes were made by adjusting the electron beam voltage. The resulting changes in the $x$ and $y$ betatron tunes gave the chromaticities. Scatter in the tune measurements was used to estimate the errors in the measurement.

The results for the chromaticities were linear with changes to the $S$ and $G$ sextupole current values, as shown in Figs. 1 and 2 for some sections through the $x$ and $y$ measurements. Altogether, chromaticity was measured at 24 locations. More points were taken in single scans for the cases where either the $S$ or $G$ currents were held constant at zero. All 24 points were used to reproduce the chromaticities with a plane described by

$$\xi_{x,y} = A + BI_S + CI_G \tag{6}$$

with $I_L = -0.145$ m$^{-3}$. The panels in Fig.~1 and 2 show the level of agreement between the plane and the $x$ and $y$ chromaticity measurements for scans along either $I_G$ or $I_S = 0$ m$^{-3}$. The quality of the agreement may be judged by the scatter of the data points about each curve. The $x$ and $y$ chromaticity planes are shown in Fig. 3.

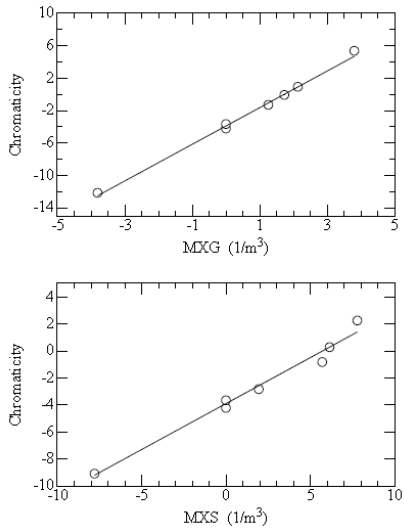

FIG 1: Measurements of the $x$ chromaticity as a function of the sextupole setting for either the MXG or MXS magnet families while the other was held at zero. The lines are the slices through the plane corresponding to the scan. The lines are sections of a plane fit to all 24 chromaticity points. The field for MXL is $-0.145$ m$^{-3}$.

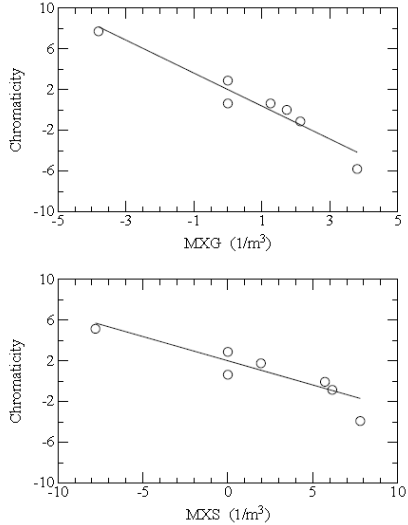

FIG 2: Measurements of the *y* chromaticity in a manner similar to Fig. 1.

FIG. 3: Values of the *x* (green) and *y* (blue) chromaticities as a function of the fields in the *S* and *G* sextupole magnet families. The planes are fits to a set of individual chromaticity measurements. The place where each plane crosses zero chromaticity is indicated by a dashed line.

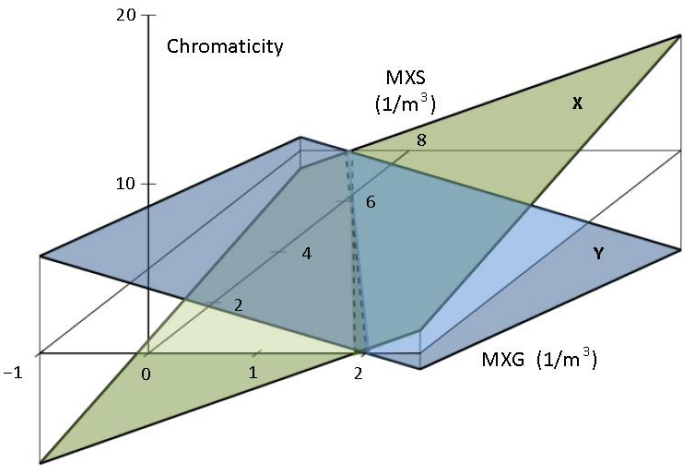

    The lines of zero chromaticity ran nearly parallel to the *SG* plane. The initial IPP lifetime scans showed little sensitivity to $I_L$, so this sextupole family was set at $-0.145$ m$^{-3}$ since this value brought the two zero lines (dashed in Fig. 3) as close together as possible. The *L* field was left at that place for the remainder of the study.

    The chromaticity results shown in Fig. 3 suggest a simpler strategy for adjusting sextupole strengths in a general 3-dimensional search: leave $I_L$ fixed and scan either $I_S$ or $I_G$ so that the scanning path crosses the zero chromaticity line. Then look at the resulting IPP lifetime values for the appearance of a maximum. Or look at the reciprocal of the lifetime for a place where it goes to zero.

    In the 2014 study we chose two beam preparations in order to be sensitive in different ways to the origins of the depolarization away from the optimum setting. The first, a "wide" beam, involved electron cooling while bunching as the initial preparation, followed by a short period of horizontal heating using white noise applied to a set of horizontal electric field plates. The white noise frequency spectrum was a band overlapping a harmonic of the betatron tune so that coupling to the beam was enhanced. This resulted in a beam that was physically wider on average than it was tall. At the time when observations of the polarization needed to begin, vertical heating was applied to the beam so that deuterons began to intersect with the front face of the carbon block

target [18]. The target limited the betatron amplitude for vertical motion, and thus limited the contribution of this degree of freedom to the beam depolarization.

The second preparation was intended to increase the spread of the longitudinal phase space in order to determine whether or not this degree of freedom mattered. The beam was first electron cooled without bunching, then, with electron cooling off, the bunching cavity was restarted. The expectation was that this would result in a beam that was physically longer ("long" beam) without having substantially more transverse motion than was there following the cooling process.

In each case, four scans were planned, as shown by bars in Fig. 4. All four scans were completed for the wide beam. A magnet cooling system failure stopped the experiment after the first three scans were completed for the long beam. Consequently our study contains only seven determinations of the optimum IPP polarization along one of these scan paths.

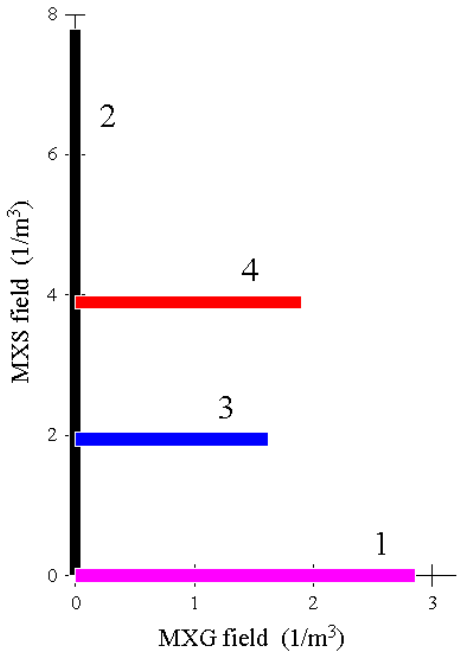

FIG. 4: In this picture of the $S \times G$ plane, four bars show the paths of four planned scans of the IPP lifetime. They are labeled "1" through "4" for later reference.

The beam stores for the wide beam were divided into several time intervals. The first 60 s were devoted to injecting, accelerating, and electron-cooling the beam. The beam was bunched throughout this time. Then cooling was turned off and for another 5 s horizontal heating white noise was applied to the beam. Beginning at 67 s, a weaker vertical white noise was applied, which drove the edges of the beam into the 17-mm thick carbon block target mounted about 3 mm above the beam center. This lasted for 10 s with the polarization vertical to allow for a measurement of the polarization. At 77 s, the rf solenoid was started, operating at the first harmonic, $(1-G\gamma)f_{\rm rev}$, of the revolution frequency. The solenoid was on for a little over 2 s, a time adjusted so that the vertical polarization went to zero. Beginning at 80 s, the data acquisition was started. At that time, some adjustment was made to the vertical position of the beam to encourage extraction of deuterons onto the carbon target. Then the window for data taking was 90 s long.

The preparation of the long beam was similar except that the horizontal heating time was used instead to start beam bunching. With cooling off at that point, the coasting beam had spread around the ring, so imposing bunching created some large amplitude synchrotron oscillations. The longer beam length was clearly visible on the beam pickups.

Each run consisted of a series of stores. There were three polarization states (positive vector, negative vector, and unpolarized), and each one in turn was used for the injected beam. The stores for each polarization state were separately accumulated and then the results added together during the data analysis. Since the purpose of the study was to produce measurements of the polarization lifetime, no effort was made to quantify the beam polarization value absolutely. What was important was knowing the starting asymmetry measured by the EDDA detector assembly for each polarization state while the polarization was still vertical. At modest beam intensities, this asymmetry was matched at 5 s after the start of the rf solenoid used to precess the polarization into the horizontal plane, indicating no significant polarization loss during rotation. These asymmetries were 0.250 for spin down and 0.229 for spin up. These were both recorded as a positive IPP since it is only the magnitude of the polarization as it rotates in the horizontal plane that is determined. On some occasions, the spin tune for a specific store was far enough away from the average for the set that there was concern the measured IPP would be reduced due to smearing of the phase during the interval of the individual time bins. In these cases, that particular store was dropped from the analysis.

Along with IPP values, detector data rates were also obtained for each time interval. In runs where the initial rate was particularly high, the measured IPP was found to be less than the typical values of 0.250 and 0.229. This led us to consider a correction to the asymmetry for rate. The larger rate came with pile-up in the signal and a background of essentially spin-independent events appearing in the data set. As the largest IPP values typically occurred for the time bin at 5 s, this bin was taken as the standard for comparison. Plotting as a function of rate in the EDDA detector, Fig. 5 shows a clear trend downward with increasing rate. So all of the measurements were corrected upward by the fraction that the curve fell below one for the rate appropriate to each time bin. This correction is particularly important since we also determined that the most reliable estimate of the IPP lifetime comes from the slope of the tangent curve fit to the data in the early part of each store. A sample of IPP data with a low rate (left panel) and a higher rate (right panel) is shown in Fig. 6.

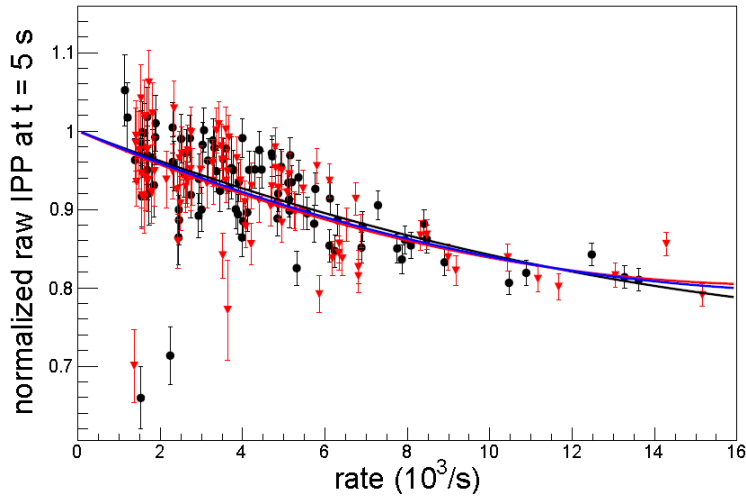

FIG. 5: The normalized value of the measured IPP in time bin 5 as a function of the detector rate. The curves are a polynomial fit to the measurements. The black curve fits the black points for spin down; the red curve fits the red points for spin up; and the blue curve fits both. The IPP values have been scaled upward by a factor for each polarization state that makes the typical low-rate bin-5 polarization equal to one. In the analysis for all time bins, the measured IPP was adjusted upward by the appropriate factor to compensate for this loss.

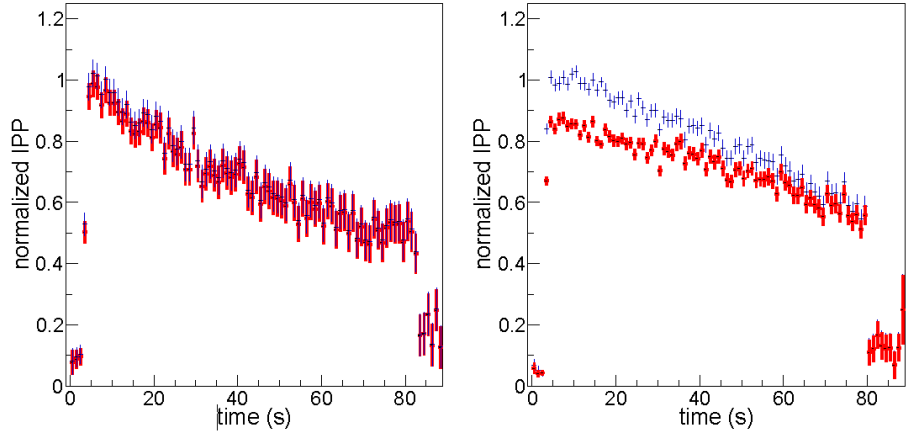

FIG. 6: Measurement of the IPP for two runs, normalized to one at t = 5 s based on the typical asymmetry values described in the text. In the left panel, the detector rate is low and not much change is apparent between red (original measurement) and blue (rate corrected measurement). The right panel contains another example for a higher rate run with a considerable initial slope change.

At the end of the experiment, all the runs were replayed. Cycles with poor information were removed from the data sample. Examples of the results for Scan 3 are shown in Fig. 7. Each one of these panels represents the results of several stores with the same polarization state and sextupole magnet settings. The average spin tune was $0.1609710 \pm 0.0000001$ among all of the runs. Rate corrections were applied.

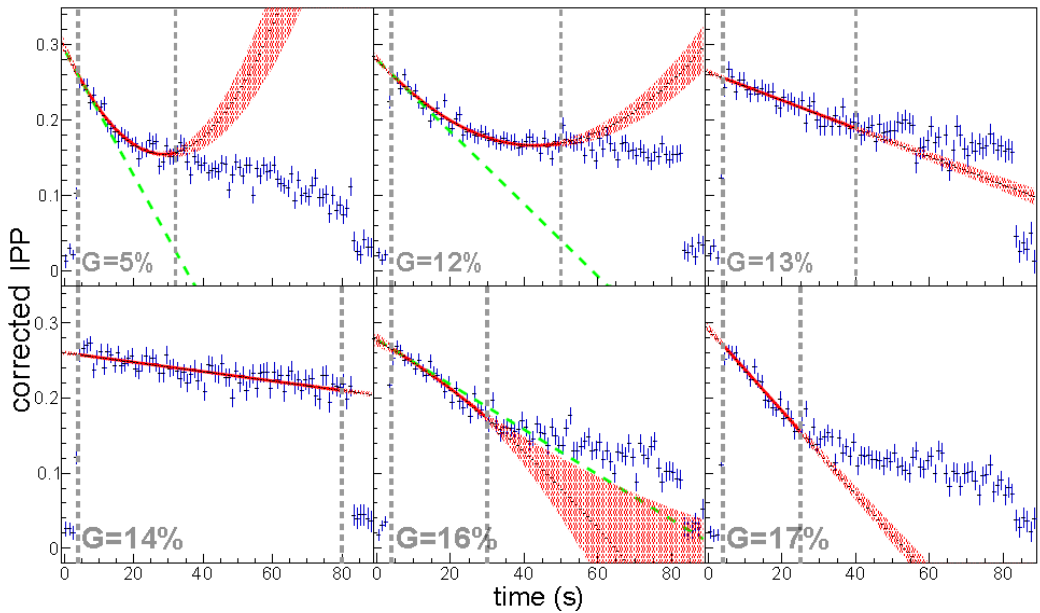

FIG. 7: Panels for representative sextupole values involved in Scan 3 of Fig. 4 for a wide beam and negative polarization. Zero time begins 80 s into the store when the data acquisition was started. The value of $I_G$ in terms of a percentage of full sextupole current are included in each panel. A region (indicated here between vertical dashed lines) near the beginning of each data set was reproduced by either a linear or a quadratic polynomial (heavy red line). The red region shows the extension of this polynomial, including statistical errors from the fitting process. The green line represents the tangent to this polynomial at 5~s where it differs from a non-linear polynomial fit.

There are various options for defining the IPP lifetime, $\tau$, based on the shape of the polarization as a function of time (see Fig. 7). An ensemble of spins with little initial spread often loses polarization slowly at first, giving a time function that is similar to a Gaussian, as seen in Fig. 1 of [1] for deuteron beams less than $10^9$ /fill. In this case the Gaussian width (half-width at 0.606 of maximum amplitude) makes sense as an IPP lifetime. This feature is present to a lesser extent in the model time dependence of Ref. [7], also created for lower beam currents. In other cases, the half-life may be appropriate (see Fig. 4 of Ref. [1]) or the width at $1/e = 0.368$ from the exponential function. The shapes of Fig. 7 present a more complicated situation (associated with beams of about $10^{10}$ /fill) without a single universal shape.

An examination of these shapes showed that the information crucial to a maximization of the IPP lifetime was often contained in the first few seconds of the time function, which could be taken from a polynomial fit to the early part of the data. In Fig. 7, a linear function did well for some time in three of the six panels. A quadratic reproduced the function in the other three panels, but resulted in two positive and one negative curvature values with no apparent pattern. Finally we chose to use either a linear or quadratic polynomial as seemed appropriate for the early part of the time function and to take only the constant and linear terms to determine the IPP lifetime, $\tau$. This lifetime was chosen to be the time over which the linear part fell to 0.606 of the initial asymmetry at t = 5 s. In each case a polynomial was fit to the measurements for as long a time as the fit reproduced the data. In some cases, a decline in polarization that began as nearly a straight line deviated within a few seconds to another course. At that point the polynomial reproduction became meaningless. Prior work with such data has indicated that the thick carbon block, which always extracts particles for the polarization measurement from the fringe of the beam, may be sampling first the halo followed by the core of the beam. With a wide beam, these two components may have different polarization distributions, resulting in some runs where there appears to be a break in the loss rate for the IPP. The time function shapes vary run to run, and may be sensitive to orbit distortions or other changes associated with the varying sextupole strengths.

In Fig. 7, the fit is shown by a red line. The red band indicates the width of the error in the line based on the statistics of the fit. The extension of the constant and linear terms in the polynomial is shown by the green line, which is visible wherever it differs from the red band. The time $\tau$ needed to reach 60.6 % of the initial asymmetry was obtained from the constant and linear polynomial coefficients. Figs. 8 and 9 show panels in which these values are represented logarithmically in a projected representation of the $S \times G$ plane.

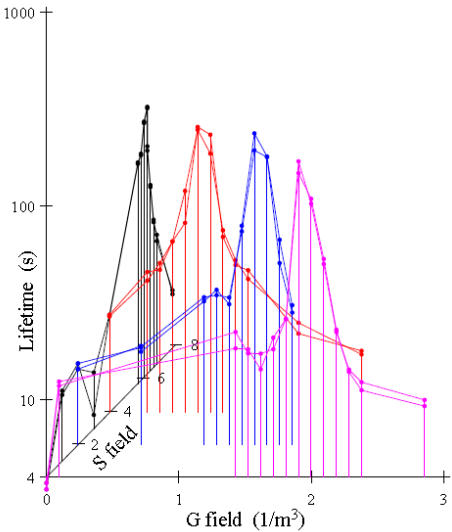

FIG. 8: Representations of the IPP lifetime $\tau$ on a projected graph as a function of $S$ and $G$ field values for the wide beam. The various scans are shown with different colors. Scan 1 is lavender, 2 is black, 3 is blue, and 4 is red. The two polarization states are plotted separately; their values overlap closely and show essentially the same information.

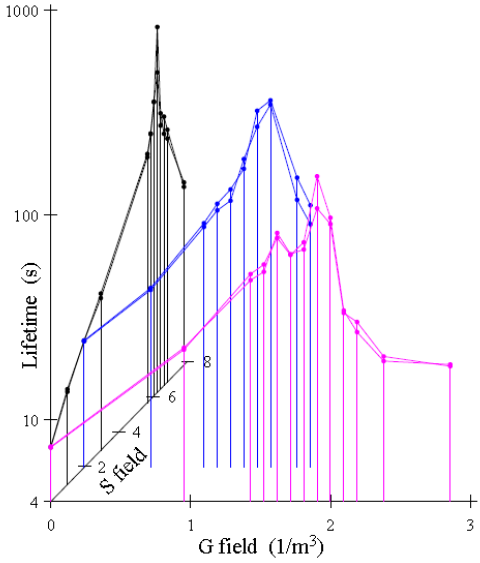

FIG. 9: Same as Fig. 8 except for the long beam. Scan 4 did not have enough data to be considered.

## IV. RESULTS AND INTERPRETATION

The curves in Fig. 8 for the wide beam show peaks that indicate the location of the longest IPP lifetimes. These peaks are more narrow than their counterparts in Fig. 9 for the long beam. This is likely a result of the deliberate horizontal heating that was applied to the beam. This spreads the distribution of betatron oscillation amplitudes and spin tunes, thus making it harder to obtain a good match for the various settings of the sextupole fields. So the IPP lifetime drops quickly away from the optimum setting. For the data in Fig. 9, no horizontal heating was applied. Instead, the beam was first cooled as a coasting beam. Afterward bunching was turned on. The resulting beam in general had a larger longitudinal extent in COSY than the wide beam. Again, there are clear peaks in the scans of IPP lifetime, but they are broader. Without the horizontal beam spreading, the lifetime is less sensitive to sextupole setting. This would suggest that the greater longitudinal extent of the beam has not introduced another source of depolarization to replace that caused by the horizontal heating. In fact, for scans 2 (black) and 3 (blue) the best IPP lifetimes are longer than for the wide beam. This suggests that, even at its best, the sextupole cancellation of the horizontal heating was not completely effective. At the same time, the higher order contribution that was expected to arrive for the long beam from $(\Delta p/p)^2$ terms has not appeared at a similar level of importance. The large differences between the lowest and highest IPP lifetimes point to the importance of making sextupole corrections in a ring where the long lifetime is important, as it is for a dedicated EDM storage ring.

The next question to consider is whether the values of the best IPP lifetimes are associated with small or zero values of the chromaticities. These locations may be obtained roughly from the peaks in Figs. 8 and 9. Earlier it was suggested that plots of $1/\tau$ from Fig. 7 versus sextupole field would have a "V" appearance as suggested by Eq. (5), especially if there was only one important contribution to the depolarization. One example from scan 2 from the wide beam runs is shown in Fig. 10.

There is a feature in the scan just above 6 m$^{-3}$ that would suggest the formation of "V" pattern. The other observation is that the errors, most of which are less than the size of the circles representing the points, are much too small to account for what appear to be fluctuations in $1/\tau$ from point to point. It is clear that, as the sextupole scans were being created, the setup of the machine itself was slowly changing over a matter of hours in addition to any response to the sextupole changes themselves. One such change might be steering of the orbit. In all cases, the sextupole values for each scan were taken in a random order to avoid the drifts appearing as trends in the measurements. But this may only cause drifts to appear as random changes from one point to the next. The decision to limit the range of the polynomial fit to only those points in Fig. 7 that could be described with a quadratic curve also served to reduce the size of the statistical errors. This problem will need to be addressed again when we attempt to estimate the errors on the location in sextupole space where the IPP lifetime is the longest. For points that are well away from the "V", the values of $1/\tau$ tend to be lower (longer IPP lifetime) than would be expected from a simple extrapolation of a linear fit to the sides of the "V". This likely indicates the growing importance of higher order corrections to the simple model of Eq. (5).

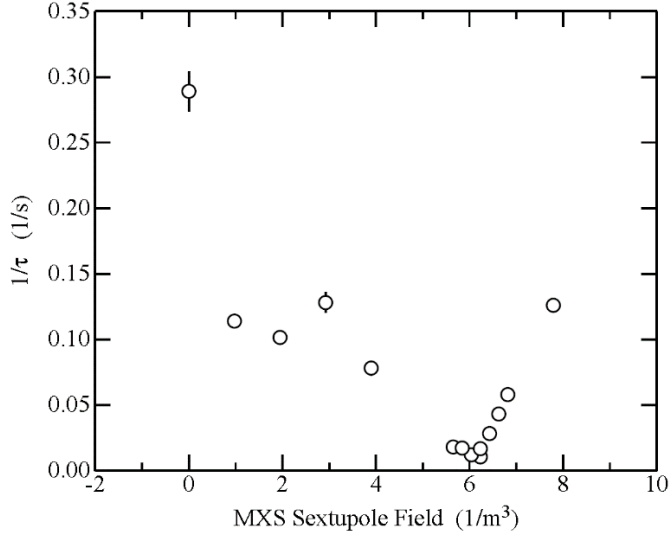

FIG. 10: Measurements of $1/\tau$ of the IPP for scan 2 and a wide beam. The value of MXG was zero. Errors shown arise only from the fit of the polynomial to the early time measurements as shown in Fig. 7. Only the points for the spin down polarization state are shown. In general, the two polarization states agree well.

In the next series of four figures, we will concentrate only on the region of the "V" for the wide beam. In order to locate more precisely the point with the best IPP lifetime, the "V" pattern may be reproduced with the absolute value of a straight line. One visualization of this fit is to reflect all of the data points and the line itself from positive to negative values at all sextupole fields that are higher than the bottom of the "V". This produces a plot similar to that of Fig. 2 in Ref. [1]. At the same time, this procedure allows all of the relevant points to be reproduced with a single (two parameter) straight-line fit. As mentioned above, some points begin to deviate from the straight line when they are away from the point of the "V". These will be excluded, as well as any points clearly having some large systematic error. The result in the area of the zero crossing for the wide beam scan 1 is shown in Fig. 11. Points outside the range of the plot were ignored in the fit. Points on the plot that were also excluded are shown with an open symbol.

It is important to note that there does not seem to be any excessive scatter in the points near the zero crossing. Points near zero that are systematically shifted away from zero might indicate other contributions to the decoherence. Such contributions must exist as the IPP lifetime cannot reach infinity, as suggested by the straight line. The quality of the fit in this region suggests that these contributions are below our present sensitivity to detect with this point spacing. The overlay of the two straight lines for the two polarization states supports the reproducibility of these results with each given run. Nevertheless, the reduced chi square value $\chi^2/\nu$ of these fits lies in the hundreds. So when the point where the lines cross zero (the position of the longest IPP lifetime) is calculated, its statistical error is enhanced by the square root of the reduced chi square value in order to compensate for systematic changes point to point. The resulting zero crossing values are given in Table I. The remaining fits for the wide beam are shown in Figs. 12, 13, and 14.

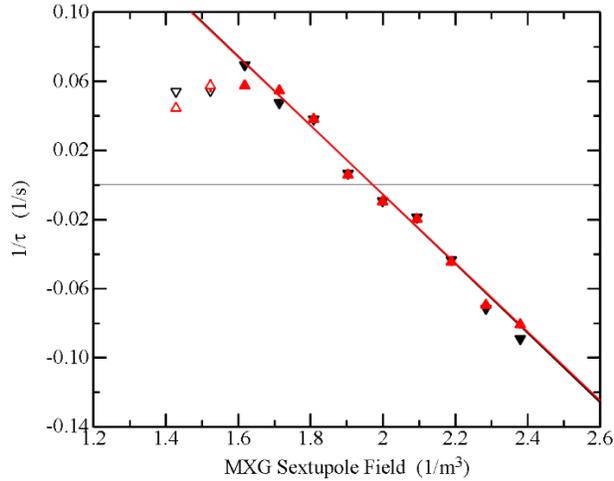

FIG. 11: $1/\tau$ points for the wide beam scan 1 as a function of the $G$ sextupole field ($S$ is zero). The black triangles are for the down polarization state and the red triangles are for the up polarization state. The filled symbols indicate those points included in the straight line fit. Points shown with open symbols were not included. Weights based on the statistical errors were used. This emphasizes points in the fit that are near the zero crossing.

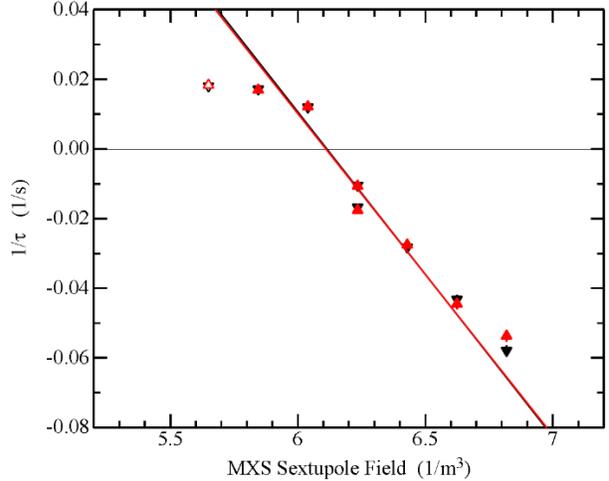

FIG. 12: Same as Fig. 11 but for wide beam scan 2, shown as a function of $S$ ($G$ is zero).

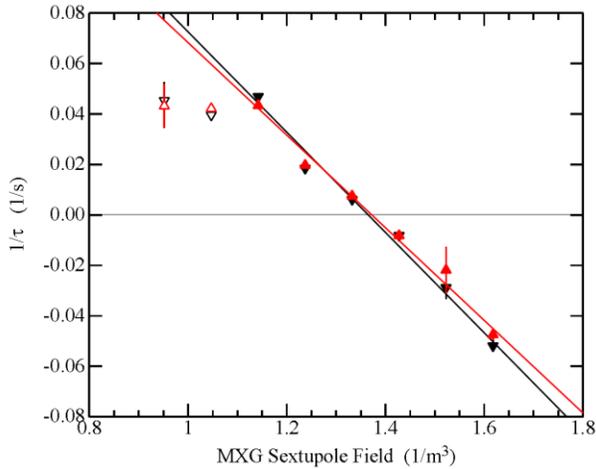

FIG. 13: Same as Fig. 11 but for wide beam scan 3, shown as a function of $G$ ($S = 1.95$ m$^{-3}$).

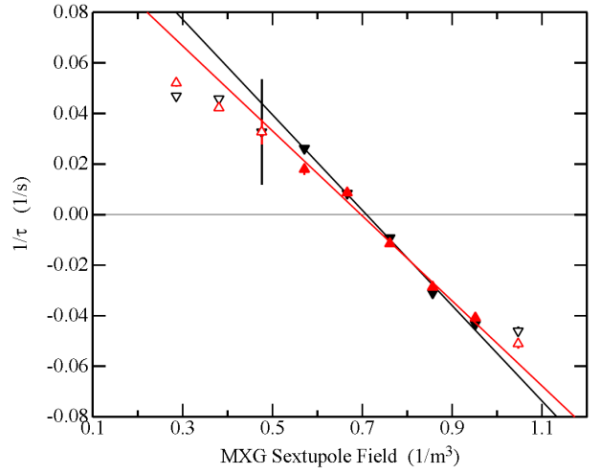

FIG. 14: Same as Fig. 11 but for wide beam scan 4, shown as a function of $G$ ($S = 3.90$ m$^{-3}$).

TABLE I. Sextupole settings for the best IPP lifetime

| Scan | Sextupole | Spin down | Spin up |
|---|---|---|---|
| Wide beam | | | |
| 1 | MXG | 1.972 ± 0.010 | 1.973 ± 0.011 |
| 2 | MXS | 6.114 ± 0.024 | 6.110 ± 0.027 |
| 3 | MXG | 1.365 ± 0.007 | 1.372 ± 0.006 |
| 4 | MXG | 0.709 ± 0.006 | 0.697 ± 0.009 |
| Long beam | | | |
| 1 | MXG | 1.93 ± 0.05 | 1.93 ± 0.05 |
| 2 | MXS | 6.23 ± 0.10 | 6.23 ± 0.10 |
| 3 | MXG | 1.53 ± 0.05 | 1.33 ± 0.05 |

Data for the long beam runs is characterized by smaller values of $1/\tau$, or longer IPP lifetimes. At the same time the "V" pattern appears to be shifted positively so that the point of the "V" no longer rests on zero. Such a shift causes a discontinuity in the straight line procedure used for the wide beam scans. A third parameter, such as the size of the shift, may be incorporated into the fitting process. But in practice the decisions about which points to include in the fit and where to locate the point of the "V" (to decide which points are to its left and right) leads to ambiguities in the results. In some cases, these are outside the bounds expected even considering the systematic deviations point to point in these measurements.

So the measurements with the long beam will be converted into an optimum sextupole setting by simple visual inspection of the data, choosing a place in a location nearest the lowest values of the slope. The error will be ± 0.05 m$^{-3}$ in $G$ and ± 0.10 m$^{-3}$ in $S$. Also, there do not appear to be any significant differences in the value taken for the two polarization states. These scans are shown in Figs. 15, 16, and 17.

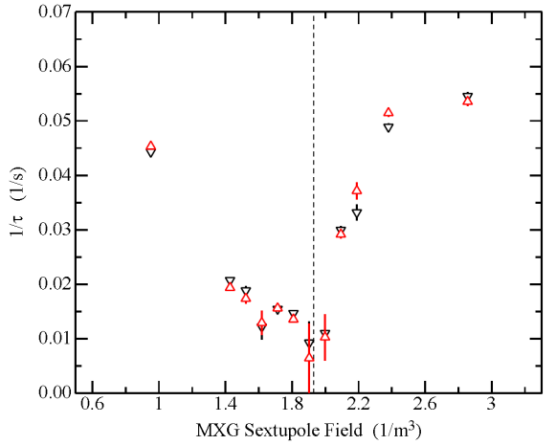

FIG. 15: Measurements made with the long beam of $1/\tau$ as a function of the $G$ sextupole field (Scan 1). The $S$ field is zero. The optimum field value is indicated by a dashed line.

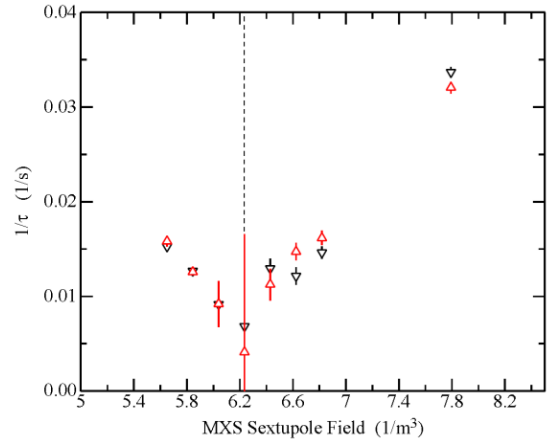

FIG. 16: Measurements made with the long beam of $1/\tau$ as a function of the $S$ sextupole field (Scan 2). The $S$ field is zero. The optimum field value is indicated by a dashed line.

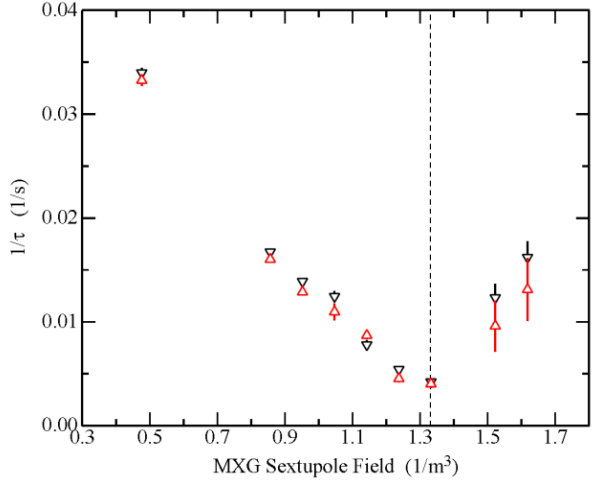

FIG. 17: Measurements made with the long beam of $1/\tau$ as a function of the G sextupole field (Scan 3). The $S$ field is 1.95 m$^{-3}$. The optimum field value is indicated by a dashed line.

The amount by which the "V" in $1/\tau$ for the long beam rises above zero is clear less than 0.01 s$^{-1}$. This could indicate the extent to which the long beam had introduced depolarization effects associated with this beam preparation method and which cannot be effectively compensated with the sextupole field manipulation. Additional tests would be needed to identify this depolarization source.

The values listed in Table I may be placed on the same template that was used for Fig. 3 of Ref. [1]. The result is Fig. 18. The two bands show where the $x$ and $y$ chromaticities were found to be zero, as indicated in Fig. 3. In this case, errors based on the internal consistency of the chromaticity measurements were used to specify bands that illustrate the one standard deviation width in the measurements of the chromaticity. The symbols used to locate the points of maximum IPP lifetime are circles for the wide beam and plus signs for the long beam. In both cases, the errors are smaller than the symbol size. The circles and plus signs are consistent with the error bands on the chromaticity, as would be expected on the basis of the formalism described in Section II.

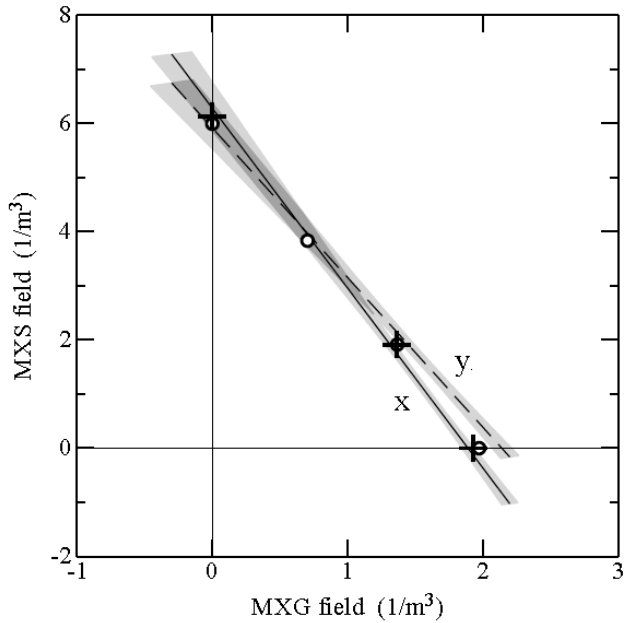

FIG. 18: Two lines with error bands show the places where the $x$ and $y$ chromaticities were consistent with zero. The locations of the points of largest IPP lifetime are shown by the circles for the wide beam and the plus signs for the long beam. For the long beam, the point from Scan 4 is missing.}

## V. CONCLUSIONS

The search for an EDM using a storage ring requires that we begin with the polarization of the beam parallel to the velocity, a situation that is unstable against momentum spread and depolarizing effects arising from betatron oscillations. In order to have a sufficient time for observation (about 1000 s) in any particular beam store, techniques are needed to suppress the depolarization of the beam once the polarization is rotated into the storage ring plane. Bunching and electron cooling together are not sufficient, so we embarked on a study to determine whether or not further progress could be made using the fields of the sextupole magnet families built into the arcs of the COSY storage ring. It was expected that the quadratic nature of the sextupole field away from the stable beam orbit could be adjusted to match the path lengthening that arises from betatron oscillations, which are still present to some extent even after electron cooling. With the right sign and scale, one would cancel the other and a major source of depolarization would be eliminated.

This proved to be correct, and a search of sextupole space on the COSY storage ring showed that there are places where the IPP lifetimes become very long. These places are not unique, but fall along a line in sextupole space. Independent measurements found that along the same line, measurements reveal almost zero chromaticity. With some sextupole field adjustment prior to running, these lines of $x$ and $y$ chromaticity may be brought together, and that combined line becomes the locus of settings with long IPP lifetimes. A simple analysis confirms this association.

The question of operating above $10^9$ deuterons/fill in the COSY experiment has to do with complications that arise at larger currents. In particular, the beam appears to separate into two components, a halo with a particularly short IPP lifetime that responds well to sextupole corrections and an inner core that does not. In addition, collective effects, such as head-to-tail oscillations of the entire beam, appear. Their impact on the polarization lifetime is not understood. These effects are likely to depend on the ring lattice used for the EDM experiment, and need to be re-evaluated in that context.

## ACKNOWLEDGEMENTS

The authors wish to thank other members of the Storage Ring EDM Collaboration [19] and the JEDI Collaboration [20] for their help with this experiment. We also wish to acknowledge the staff of COSY for providing good working conditions and for their support of the technical aspects of this experiment. This work has been financially supported by the Forschungszentrum Jülich via COSY-FFE, by the European Research Council Advanced-Grant (srEDM, No. 6984340) of the European Union, and by a grant from the Shota Rustaveli National Science Foundation of the Republic of Georgia (SRNSF grant No. 217854, "A first-ever measurement of the EDM of the deuteron at COSY"). One of us (G.G.) acknowledges the support of a travel grant (FONDI 5 X 1000 ANNO 2011) from the University of Ferrara. One of us (A.S.) is supported by a grant from the Russian Science Foundation (Grant No.\ RNF-16-12-10151).

1. G. Guidoboni *et al.*, Phys. Rev. Lett. **117**, 054801 (2016).
2. R. Maier, Nucl. Instrum. Methods Phys. Res. A **390**, 1 (1997).
3. V. Anastassopoulos *et al.*, Rev. Sci. Instrum. **87**, 115116 (2016).
4. M. Rosenthal, Ph.D. thesis, RWTH Aachen University, available at http://collaborations.fz-juelich.de/ikp/jedi/public_files/theses/Thesis_Rosenthal.pdf .
5. F.J.M. Farley, K. Jungmann, J.P. Miller, W.M. Morse, V.F. Orlov, B.L. Roberts, Y.K. Semertzidis, A. Silenko, and E.J. Stephenson, Phys. Rev. Lett. **93**, 052001 (2004).
6. Yu. Senichev, R. Maier, D. Zyuzin, and N. Kylabukova, *Proc. of the Fourth Int. Part. Accel. Conf.,* Shanghai, China, edited by Z. Dai, C. Petit-Jean-Genaz, V.R.W. Schaa, and C. Zhang, WEPE036, p. 2579.
7. Z. Bagdasarian *et al.*, Phys. Rev. ST Accel. Beams **17**, 052803 (2014).
8. S.Y. Lee, *Accelerator Physics* (World Scientific, Singapore, 1999).
9. A.E. Chao and M. Tigner, *Handbook of Accelerator Physics and Engineering* (World Scientific, Singapore, 1998).
10. Y. Shoji, Phys. Rev. ST Accel. Beams **8**, 094001 (2005).
11. I.A. Koop and J.M. Shatunov, *Proc. of the First European Particle Accelerator Conf.*, Rome, 1998, edited by S. Tazzari (World Scientific, Singapore, 1989), p. 738.
12. S. Serednyakov, V.A. Sidorov, A.N. Skrinsky, G.M. Tumaikim, and J.M. Shatunov, Phys. Lett. **66B**, 102 (1977).
13. I. Vassermann *et al.*, Phys. Lett. B **187**, 172 (1987).
14. P. Benati *et al.*, Phys. Rev. ST Accel. Beams **15**, 124202 (2012); erratum ibid. **16**, 049901 (2013).
15. E. Weise, Ph.D. thesis, University of Bonn, 2000, http://edda.hiskp.uni-bonn,de/dipldiss.html .
16. D. Albers *et al.*, Eur. Phys. J. A **22**, 125 (2004).
17. J. Bisplinghoff *et al.*, Nucl. Instrum. Methods Phys. Res., Sect. A **329**, 151 (1993).
18. N.P.M. Brantjes *et al.*, Nucl. Instrum. Methods Phys. Res. A **664**, 49 (2012).
19. http://www.bnl.gov/edm/ .
20. http://collaborations.fz-juelich.de/ikp/jedi .